\documentclass[aps,pre,showpacs,showkeys,12pt,a4paper,amsmath,amssymb]
{revtex4-1}
\usepackage[english]{babel}
\usepackage{graphicx}

\renewcommand{\vec}[1]{\ensuremath{\mathbf{#1}}}

\large
\providecommand{\figref}[1]{Fig.~\ref{#1}}

\newcommand{\murm}{{\usefont{U}{psy}{m}{n}m}}

\begin{document}

\title{Three-dimensional Coherent Structures of Electrokinetic Instability}

\author{\firstname{E.~A.}~\surname{Demekhin}}
\email{edemekhi@gmail.com}
\affiliation{Department of Computation Mathematics and Computer Science,
Kuban State University, Krasnodar, 350040, Russian Federation}
\affiliation{Laboratory of General Aeromechanics, Institute of Mechanics,
Moscow State University, Moscow, 117192, Russian Federation}

\author{\firstname{N.~V.}~\surname{Nikitin}}
\affiliation{Laboratory of General Aeromechanics, Institute of Mechanics,
Moscow State University, Moscow, 117192, Russian Federation}

\author{\firstname{V.~S.}~\surname{Shelistov}}
\affiliation{Scientific Research Department,
Kuban State University, Krasnodar, 350040, Russian Federation}

\begin{abstract}
A direct numerical simulation of the three-dimensional elektrokinetic
instability near a charge selective surface (electric membrane, electrode, or
system of micro-/nanochannels) is carried out and analyzed. A~special
finite-difference method was used for the space discretization along with a
semi-implicit $3\frac{1}{3}$\nobreakdash-step Runge--Kutta scheme for the
integration in time. The calculations employed parallel computing. Three
characteristic patterns, which correspond to the overlimiting currents, are
observed: (a)~two-dimensional electroconvective rolls, (b)~three-dimensional
regular hexagonal structures, and (c)~three-dimensional structures of
spatiotemporal chaos, which are a combination of unsteady hexagons, quadrangles
and triangles. The transition from (b) to (c) is accompanied by the generation
of interacting two-dimensional solitary pulses.
\end{abstract}



\typeout{WARNING: page enlarged}	\enlargethispage{3em}
\maketitle

\section*{Introduction}

Problems of electrokinetics and micro-/nanofluidics have recently attracted a
great deal of attention due to rapid developments in micro-, nano-, and
biotechnology. Among the numerous modern micro-/nanofluidic applications of
electrokinetics are micropumps, micromixers, {\murm}TAs, desalination, fuel cells,
etc. (see \cite{ChangDem}). The fundamental interest in the problem is connected
with a novel type of electrohydrodynamic instability at the micro- and
nanoscales: the electrokinetic instability. The electrokinetic instability
describes the generation of nonlinear coherent structures near a
charge-selective surface under a drop in the electric potential. This
instability was recently theoretically predicted by Rubinstein and Zaltzman
\cite{Rub3,Rub3a,Rub10} and experimentally confirmed in \cite{Rub23,Mal,Rub12,
RubRub1,YCh,KmHn}. The linear stability theory of the 1D quiescent steady-state
solution, based on a systematic asymptotic analysis of the problem, was
developed by Zaltzman and Rubinstein \cite{Rub13}. Different aspects of the
linear stability of the one-dimensional (1D) solution were also studied in
\cite{Dem,DemSh,KD2}.

Not all important facts of the electrokinetic instability can be described by
the asymptotic analysis and linear stability theory. Only direct numerical
simulations (DNS) of the Nernst--Planck--Poisson--Navier--Stokes equations give
reliable tools to study all the details of the electrokinetic instability. In
the first DNS studies \cite{DemShel,ShelDem,DemChang,Han1, Mani,DemNikSh}, the
nontrivial stages of noise-driven nonlinear evolution towards overlimiting
regimes were identified, a space charge in the extended space charge region was
found to have a typical spike-like distribution, the dynamics of the spikes was
investigated along with the physical mechanisms of the secondary instabilities,
and it was demonstrated that the transition between the limiting and the
overlimiting current regimes can exhibit a hysteretic behavior (subcritical
bifurcation). 

All these results were obtained in the two-dimensional (2D) formulation.
Actually, the electrokinetic instability is three-dimensional (3D) and such a
fact can affect the DNS results. In the present paper, for the first time, 3D
numerical simulations of the electrokinetic instability are carried out.
White-noise initial conditions to mimic ``room disturbances'' and the subsequent
natural evolution of the solution are considered.

The following regimes, which replace each other as the potential drop between
the selective surfaces increases, are obtained: a 1D quiescent steady-state
solution, 2D steady electroconvective rolls (vortices), unsteady 2D vortices
regularly or chaotically changing their parameters, steady 3D hexagonal
patterns, and a chaotic spatiotemporal 3D motion.

The space-charge region profile for 2D rolls has long flat and short wedge-like
regions with a cusp at the top. The cusp angle does not depend on the parameters
and is about $111^\circ$. The 3D hexagonal structures consist of six wedge-like
lateral faces and six pyramids are located at their intersection. The angle of
wedge-like faces is close to that for the 2D rolls, and the dependence of this
angle on the parameters of the problem is also weak. A rough evaluation of the
pyramid angle gives its value as about $87^\circ$.

An interesting phenomenon found is the generation of two-dimensional running
solitary waves either inside the hexagonal structure or at one of its lateral
sides. If another solitary wave forms, a complex head-on or an oblique
pulse--pulse interaction occurs. For large drop of potential, the pulse--pulse
interaction becomes strong enough to destroy the hexagonal structure and a
transition to spatiotemporal chaos results from such strong pulse interaction.

\section*{Formulation of the problem}

A~symmetric, binary electrolyte with a diffusivity of cations and anions
$\tilde{D}$, dynamic viscosity $\tilde{\mu}$, and electric permittivity
$\tilde{\varepsilon}$, and bounded by ideal, semiselective ion-exchange membrane
surfaces, $\tilde{y}=0$ and $\tilde{y}=\tilde{L}$, is considered. Notations with
tildes are used for the dimensional variables, as opposed to their dimensionless
counterparts without tildes. $\tilde{x}$, $\tilde{y}$ and $\tilde{z}$ are the
coordinates, where $\tilde{x}$ and $\tilde{z}$ are directed along the membrane
surface, and $\tilde{y}$ is normal to it.

The characteristic quantities to make the system dimensionless are as follows:
$\tilde{L}$ is the distance between the membranes; $\tilde{L}^2/\tilde{D}$ is
the characteristic time; the dynamic viscosity $\tilde{\mu}$ is taken as a
characteristic dynamical quantity; the thermic potential $\tilde{\Phi}_0=
\tilde{R}\tilde{T}/\tilde{F}$ is taken as a characteristic potential; and the
bulk ion concentration at the initial time $\tilde{c}_0$ is the characteristic
concentration. Here, $\tilde{F}$ is Faraday's constant, $\tilde{R}$ is the
universal gas constant, and $\tilde{T}$ is the absolute temperature.

The electroconvection is described by the equations for ion transport, Poisson's
equation for the electric potential, and the Stokes equations for a creeping
flow:
\begin{align}
\frac{\partial c^\pm}{\partial t} + \vec{u}\cdot\nabla c^\pm &=
 \pm\nabla\cdot\left(c^\pm\nabla\Phi\right) + \nabla^2 c^\pm,
 & \nu^2\nabla^2\Phi &= c^--c^+ \quad \equiv -\rho,	\label{eq1}\\
-\nabla\Pi + \nabla^2\vec{u} &= \frac{\varkappa}{\nu^2}\rho\nabla\Phi,
 & \nabla\cdot\vec{u} &= 0.	\label{eq3}
\end{align}
Here, $c^\pm$ are the concentrations of the cations and anions;
$\vec{u}=\{u,v,w\}$ is the fluid velocity vector; $\Phi$ is the electrical
potential; $\Pi$ is the pressure; $\nu$ is the dimensionless Debye length or
Debye number,
\[
\nu = \frac{\tilde\lambda_D}{\tilde{L}}, \qquad \tilde\lambda_D = 
 \left(\frac{\tilde\varepsilon\tilde\Phi_0}{\tilde{F}\tilde{c}_0}\right)^{1/2}
 = \left(\frac{\tilde\varepsilon\tilde{R}\tilde{T}}{\tilde{F}^2\tilde{c}_0}
 \right)^{1/2},
\]
and $\varkappa=\tilde\varepsilon\tilde\Phi_0^2/\tilde\mu\tilde{D}$ is a coupling
coefficient between the hydrodynamics and the electrostatics. It characterizes
the physical properties of the electrolyte solution and is fixed for a given
liquid and electrolyte.

This system of dimensional equations is complemented by the proper conditions at
the lower and upper boundaries, $y=0$ and $y=1$:
\begin{equation}\label{eqq5}
c^+ = p, \qquad
 -c^-\frac{\partial\Phi}{\partial y} + \frac{\partial c^-}{\partial y} = 0,
 \qquad \vec{u} = 0.
\end{equation}
The potential drop between the membranes is $\Delta V$.

The first boundary condition, prescribing an interface concentration equal to
that of the fixed charges inside the membrane, is asymptotically valid for large
$p$ and was first introduced by Rubinstein, see, for example,
\cite{Rub13}. The second boundary condition means there is no flux for negative
ions, and the last condition is that the velocity vanishes at the rigid surface.
The spatial domain is assumed to be infinitely large in the $x$- and
$z$-directions, and the boundedness of the solution as $x,z\to\pm\infty$ is
imposed as a condition.

Adding initial conditions for the cations and anions completes the formulation
\eqref{eq1}--\eqref{eqq5}. These initial conditions arise from the following
viewpoint: when there is no potential difference between the membranes, the
distribution of ions is homogeneous and neutral. This corresponds to the
condition $c^+ = c^- = 1$. Some kind of perturbation should be superimposed on
this distribution, which is natural from the viewpoint of experiment. The
so-called ``room disturbances'' determining the external low-amplitude and
broadband white noise should be imposed on the concentration:
\begin{equation}\label{eqq67}
t=0: \qquad c^{\pm} = 1 +\int\limits_{-\infty}^{+\infty} \hat{c}^{\pm}(k,m)
 e^{-i(kx+mz)} \, dk dm.
\end{equation}
Here, the phase of the complex function $ \hat{c}^{\pm}(k,m)$ is assumed to be a
random number with a uniform distribution inside the interval $[0,2\pi]$.

The characteristic electric current $j$ at the membrane's surface is referred to
the limiting current, $j_{\lim}=4$,
\begin{equation}\label{eq777}
j = \frac{1}{4}\left(c^+\frac{\partial\Phi}{\partial y} +
 \frac{\partial c^+}{\partial y}\right) \text{ for } y = 0.
\end{equation}
It is also convenient for our further analysis to introduce the electric current
averaged with respect to the membrane{'s} surface $l_x\times l_z$ and to the
time:
\begin{equation}\label{eq76}
\langle j(t) \rangle = \frac{1}{{l_xl_z}}
 \int\limits_0^{l_x}{\int\limits_0^{l_z}} j(x,z,t) \, dx dz,
 \qquad J = \lim_{T\to\infty} \frac{1}{T} \int\limits_0^T
 \langle j(t)\rangle \, dt.
\end{equation}

The problem is characterized by three dimensionless parameters: $\Delta V$,
$\nu$ (which is a small parameter), and~$\varkappa$. The dependence on the
concentration, $p$, for the overlimiting regimes is practically absent, and thus
$p$ is not included in the mentioned parameters: in all calculations, $p = 5$.

The problem is solved for $\nu = 10^{-3}$, $\varkappa = 0.05 \div 0.5$, and the
dimensionless potential drop varied within $\Delta V = 0 \div 60$.

\section*{The numerical method}

The numerical approach of \cite{DemNikSh} is generalized for the solution of the
system \eqref{eq1}--\eqref{eq76}. A~finite-difference method with second-order
accuracy is applied for the spatial discretization. A~uniform grid is used in
the homogeneous tangential $x$- and $z$-directions; the grid is stretched in
the normal $y$-direction via a $\tanh$ stretching function in order to properly
resolve the thin double-ion layers attached to the membrane surfaces. When a
fine spatial resolution is used, our system represents a stiff problem. In order 
to solve this problem, implicit methods require the inversion of rather large
matrices and thus are extremely costly, while explicit methods of time
advancement require a very small time-step and, hence, are prohibitively
ineffective. A semi-implicit method is found to be a reasonable compromise: only
a part of the right-hand side of the system is treated implicitly \cite{Niki1}.
A~special semi-implicit $3\frac{1}{3}$\nobreakdash-step Runge--Kutta scheme is
used for the eventual integration in time. The details of the numerical scheme
will be presented elsewhere \cite{DemNikSh2}. 

For the natural ``room disturbances,'' the infinite spatial domain is changed to
a large enough finite domain that has lengths $l_x=l_z=l$ in both spatial
dimensions, with the corresponding wave number $k=2\pi/l$. The condition that
the solution as $x,z \to \infty$ be bounded is changed to periodic boundary
conditions. The length of the domain $l$ has to be taken large enough to make
the solution independent of the domain size. The wave number $k$ is taken to be
1.

The parallel computing was carried out at the supercomputer ``Chebyshev'' of the
computer cluster SKIF of the Moscow State University, using up to 256 MPI
processors. A~resolution of 256 points in the $x$- and $z$-directions along with
128 points in the $y$-direction provides adequate results. In order to check
their accuracy, the number of points in all directions for some simulations was
doubled.

\section*{Simulation results}

The system \eqref{eq1}--\eqref{eq76} has a 1D quiescent steady state solution
which describes the underlimiting and limiting currents. For the limiting
currents in a small vicinity of the charge selective surface, there is a thin
electric double-ion layer (EDL); right away from this surface an equilibrium
diffusion layer forms; the volt--current (VC) curve obeys a linear Ohmic
relation. For the limiting currents, the VC curve has a typical saturation of
the electric current with respect to the drop of potential. In order to explain
this behavior, Rubinstein and Shtilman \cite{Rub1} came up with the idea of the
nonequilibrium nature of the EDL and of the extended space charge (ESC) region,
$0<y<y_m$, which is much thicker than the EDL. For the underlimiting and
limiting currents, the diffusion is balanced by electromigration, there is no
contribution of convection to the ion flux, and the ESC layer thickness, $y_m$,
is uniform along the membrane surface. (The boundary of the ESC region, $y_m$,
gives a convenient value to describe the electrokinetic patterns. This boundary
is a conventional value; we define it by taking 5\% of the maximal value of the
space charge density in the ESC region).

The appearance of the extended space charge for the limiting current regimes
leads for $\Delta V>\Delta V_*$ to a special kind of electrohydrodynamic
instability, the electrokinetic instability (see \cite{Rub3,Rub3a,Rub13}).
A~small inhomogeneity in $y_m(x,z)$ along the membrane results in a convective
motion of the fluid in the inner ESC region with a tangential slip velocity and
leads to the growth of the perturbations: the 1D steady-state equilibria lose
their stability and overlimiting currents eventually arise. For the overlimiting
currents, the third mechanism, convection, significantly contributes to the ion
flux. 

\begin{figure}[hbtp]
\centering
\includegraphics[angle=270,width=\textwidth]{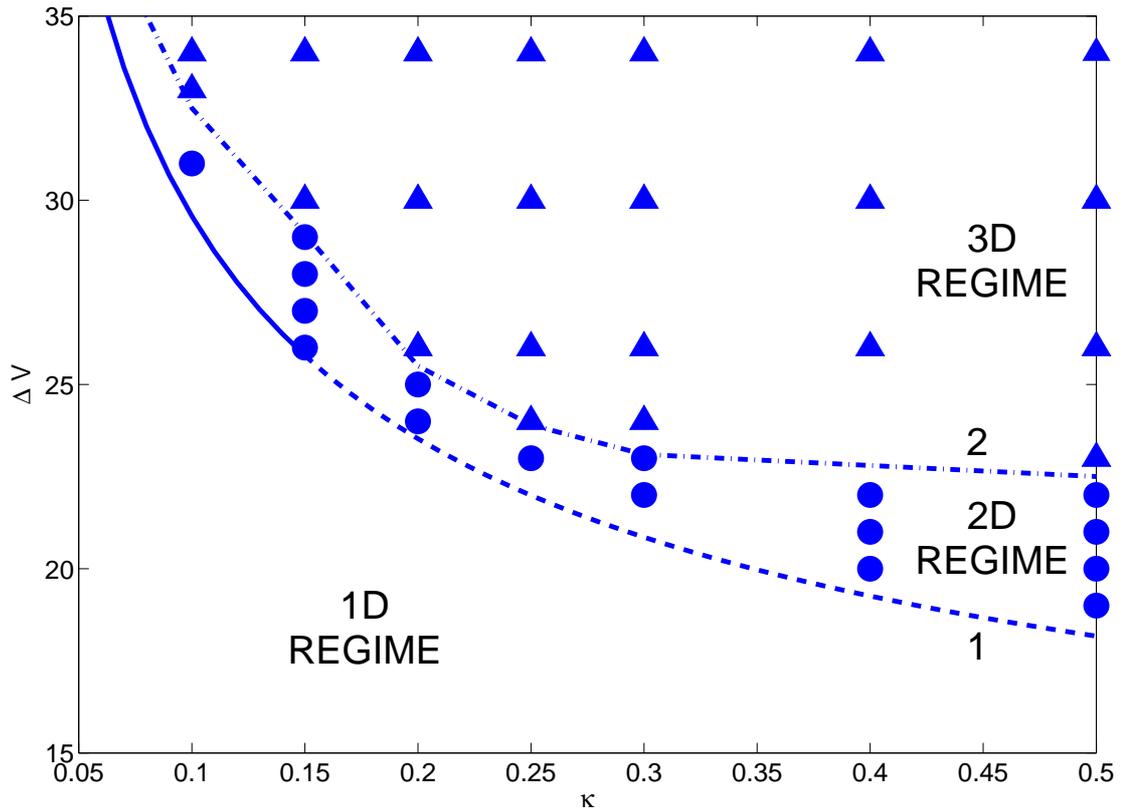}
\caption{(Color online) Map of regimes and bifurcations, $\Delta V$ vs.
$\varkappa$ for $\nu=10^{-3}$. The solid line is the neutral stability curve
\cite{DemNikSh} which separates the 1D and 2D regimes. Circles correspond to
realizations of 2D regimes; triangles, to realization of 3D regimes, so that the
2D--3D transition happens on the dashed line.}
\label{fig:2d3d}
\end{figure}

In \figref{fig:2d3d} a map of the regimes and bifurcations is presented: the
first coordinate is the potential drop $\Delta V$ and the other is the coupling
coefficient~$\varkappa$. The curve~1 corresponds to the threshold of
instability: for $\varkappa<0.151$, the bifurcation is supercritical and this
part of~1 is pictured by the solid line; for $\varkappa > 0.151$, the
bifurcation is subcritical, this part of the curve is shown by the dashed line
(see \cite{DemNikSh}). The 1D regimes and the limiting currents are located
below~1. The circles and triangles stand for the 2D and 3D regimes,
respectively. The dash-dot line~2 separates these regimes and corresponds to the
2D--3D transition. 

\begin{figure}[hbtp]
\centering
\includegraphics[angle=270,width=\textwidth]{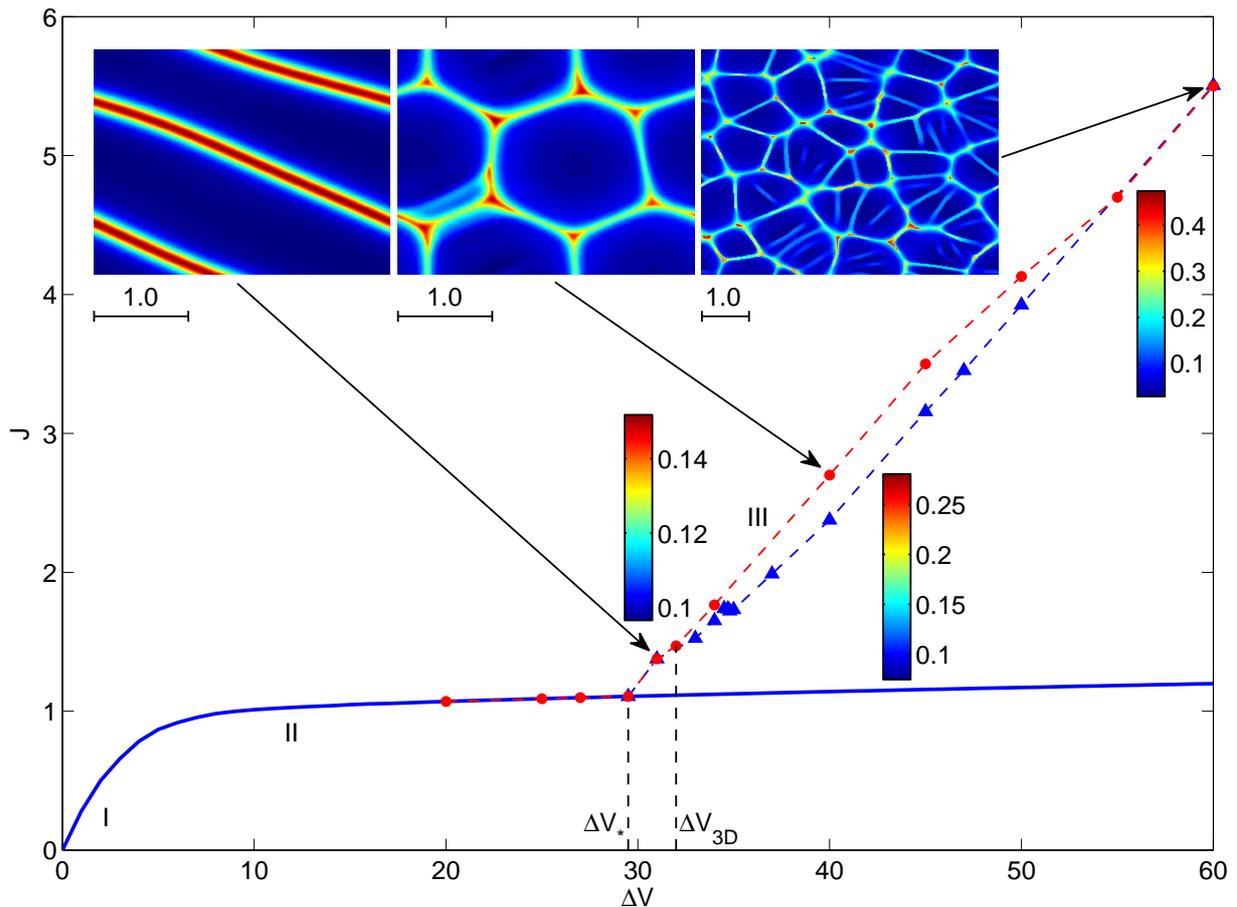}
\caption{(Color online) VC characteristics for $\varkappa=0.1$ and
$\nu=10^{-3}$: I, II and III stand for underlimiting, limiting, and overlimiting
currents, respectively. The dashed line joining the circles corresponds to our
3D simulations; the dashed line joining the triangles corresponds to 2D
simulations. The inset shows three typical electrokinetic patterns: 2D
electroconvective rolls, regular 3D hexagonal structures, and chaotic 3D
structures.}
\label{fig:vc}
\end{figure}

The dependence of the average of the electric current $J$ over the membrane
surface $l_xl_z=l^2$ and the elapsed time $t$ (see Eq.\eqref{eq777}) on the
potential drop $\Delta V$ is a convenient integral characteristic of the
regimes. Such a VC dependence is shown in \figref{fig:vc} for a typical value of
the coupling coefficient $\varkappa=0.1$, where the bifurcation is
supercritical. Portions of the VC dependence, I, II and III, stand for the
underlimiting, limiting, and overlimiting currents, respectively. The dashed
line in \figref{fig:vc} joining the circles corresponds to 3D simulations. We
find it instructive to plot in this figure also the results of the 2D
simulations; they are shown by the dashed line joining the triangles
corresponding to the 2D simulations. Until the point $\Delta V=
\Delta V_\text{3D}$, both dependences coincide. This means that for a small
supercriticality, the electrokinetic instability is two-dimensional. This points
to the fact that 3D effects increase the ion flux in comparison with the
two-dimensional regime, but this increase is not large, about $5\% \div 10\%$.
Moreover, for large enough $\Delta V$, $\Delta V>55$, this difference
practically disappears.

Our simulations show that four basic coherent structures can be found during the
evolution: 2D electroconvective rolls (vortices), squares, triangles, and
hexagons.
\begin{enumerate}\renewcommand{\labelenumi}{(\alph{enumi})}
\item The first coherent structure, spatially periodic stationary
electroconvective rolls, can be realized as an attractor, as $t\to\infty$, only
in a narrow band near the threshold of instability, between curves $1$ and
$2$ of \figref{fig:2d3d}. This is reminiscent of the Rayleigh-B\'enard
convection (see \cite{CrHog}) when near the threshold hexagons and squares are
unstable to rolls and there is a closed region of their stability called the
``Busse balloon.'' Note that the bifurcation picture is different for the
B\'enard--Marangoni convection, where stable hexagonal patterns can arise at the
threshold (see \cite{Nep}). The two dimensional coherent structures are of
particular interest because of the relative simplicity of their investigation in
the 2D formulation. These solutions were analyzed in detail in
\cite{DemShel,ShelDem,DemChang,Han1,Mani,DemNikSh}. 
\item For $\Delta V > \Delta V_\text{3D}$, above line~2 of \figref{fig:2d3d},
the 2D electroconvective rolls become unstable to three-dimensional
perturbations. Theoretically there are three candidates to inherit stability and
be a new attractor: squares, triangles, and hexagons \cite{CrHog}. Our
simulations show that for the electrokinetic instability, steady and regular
squares and triangles do not exist: they can be seen during the evolution only
as a transitional state.
\item Regular steady-state hexagonal patterns are formed just above line~2 of
\figref{fig:2d3d}. The white noise initial perturbations eventually evolve
towards steady hexagonal patterns.
\item As the driving $\Delta V-\Delta V_*$ is increased, the ordered hexagonal
structures break down to complex and highly disordered states and the behavior
becomes chaotic in time and space. The flow is a combination of unsteady
hexagons, quadrangles and triangles.
\end{enumerate}

To complete the VC dependence, three characteristic electrokinetic patterns are
shown in the inset to \figref{fig:vc}: 2D electroconvective vortices, regular 3D
hexagonal structures, and chaotic 3D structures (combinations of unsteady
hexagons, quadrangles and triangles). The arrows show the place of these
structures at the VC curve and a typical potential drop for their realization.
A~movie of the evolution of these structures can be found in \cite{Vis}.

\begin{figure}[hbtp]
\raggedright
\includegraphics[origin=rb,angle=270,width=.6\textwidth]{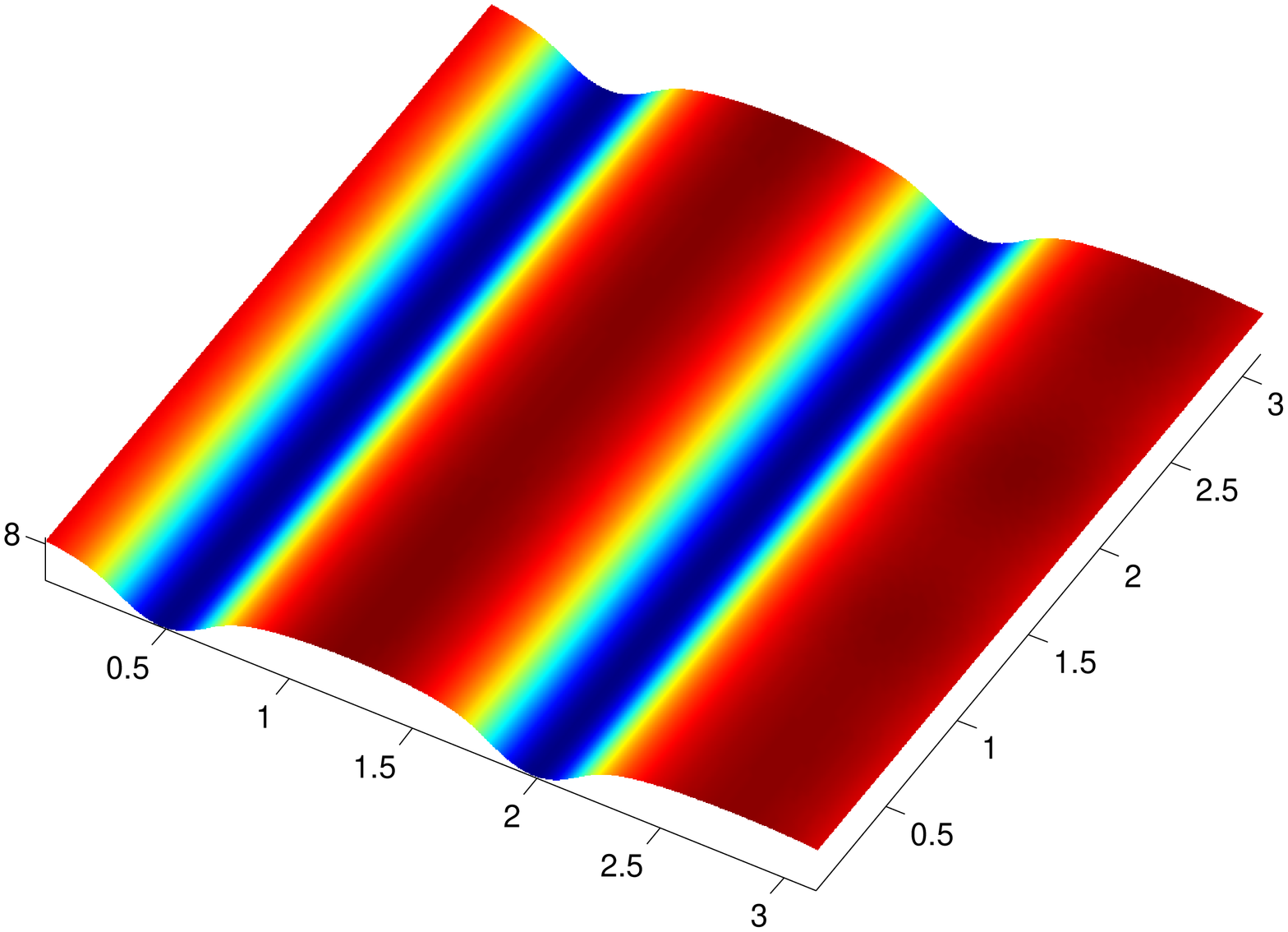}

\centering
\includegraphics[width=.6\textwidth]{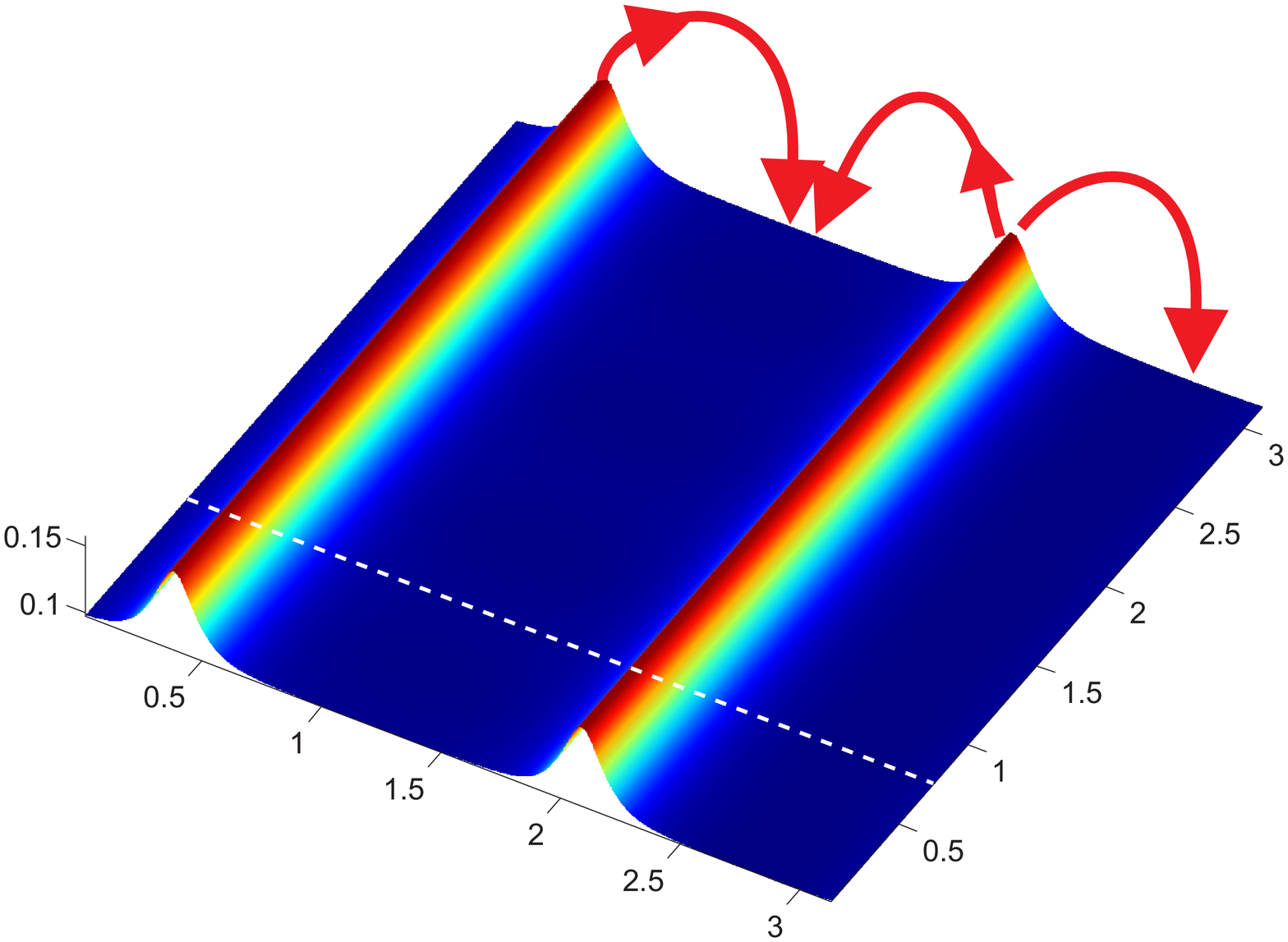}
\begin{picture}(0,0)
\put(-240,230){$j(x,z)$}
\put(-240,10){$y_m(x,z)$}
\put(-50,10){$\rho(x,y,z_*)$}
\end{picture}
\includegraphics[width=.38\textwidth]{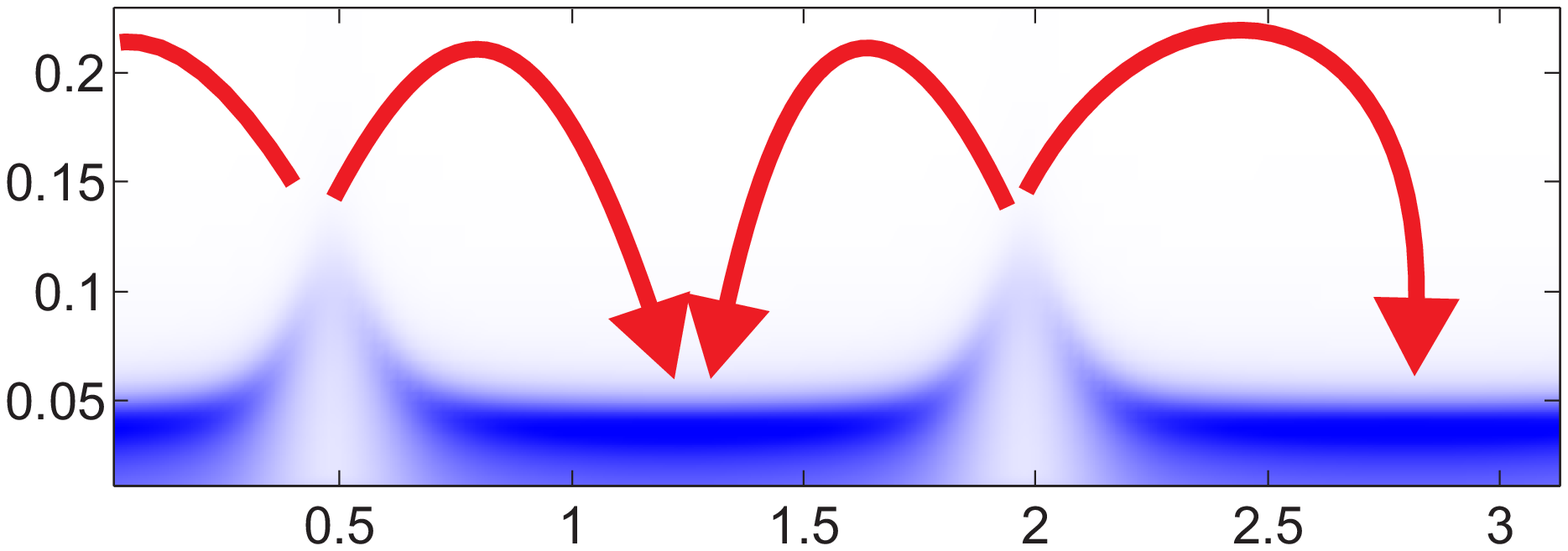}
\caption{(Color online) Electroconvective rolls for $\Delta V=33$,
$\varkappa=0.1$, and $\nu=10^{-3}$. Profile of $y_m(x,z)$, cross-section of
charge density $\rho$ at $z=z_*$, and distribution of the electric current
$j(x,z)$. Vortex pairs of liquid flow are shown schematically by the arrows.}
\label{fig:2d}
\end{figure}

\begin{figure}[hbtp]
\raggedright
\includegraphics[origin=rb,angle=270,width=.6\textwidth]{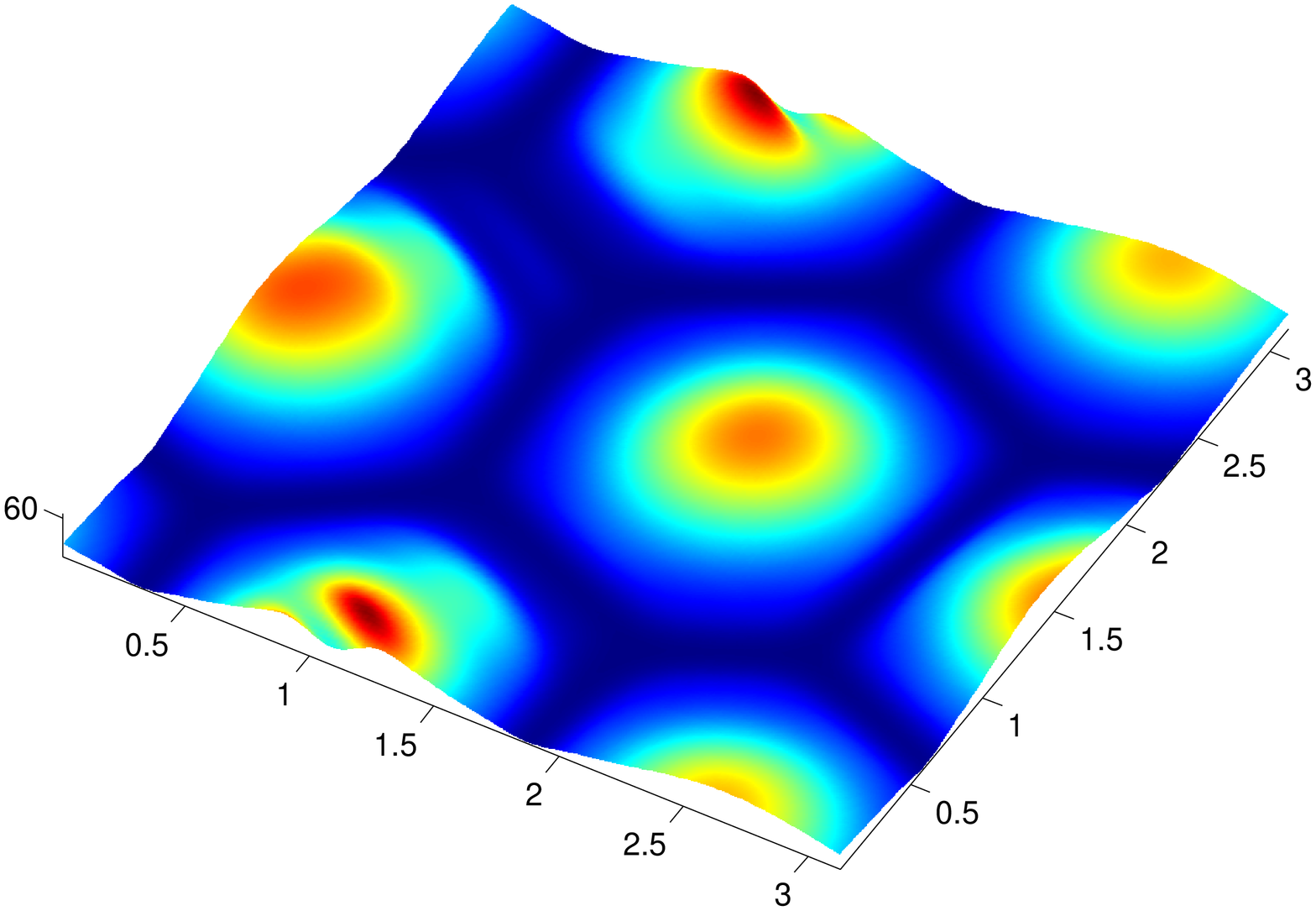}

\centering
\includegraphics[width=.6\textwidth]{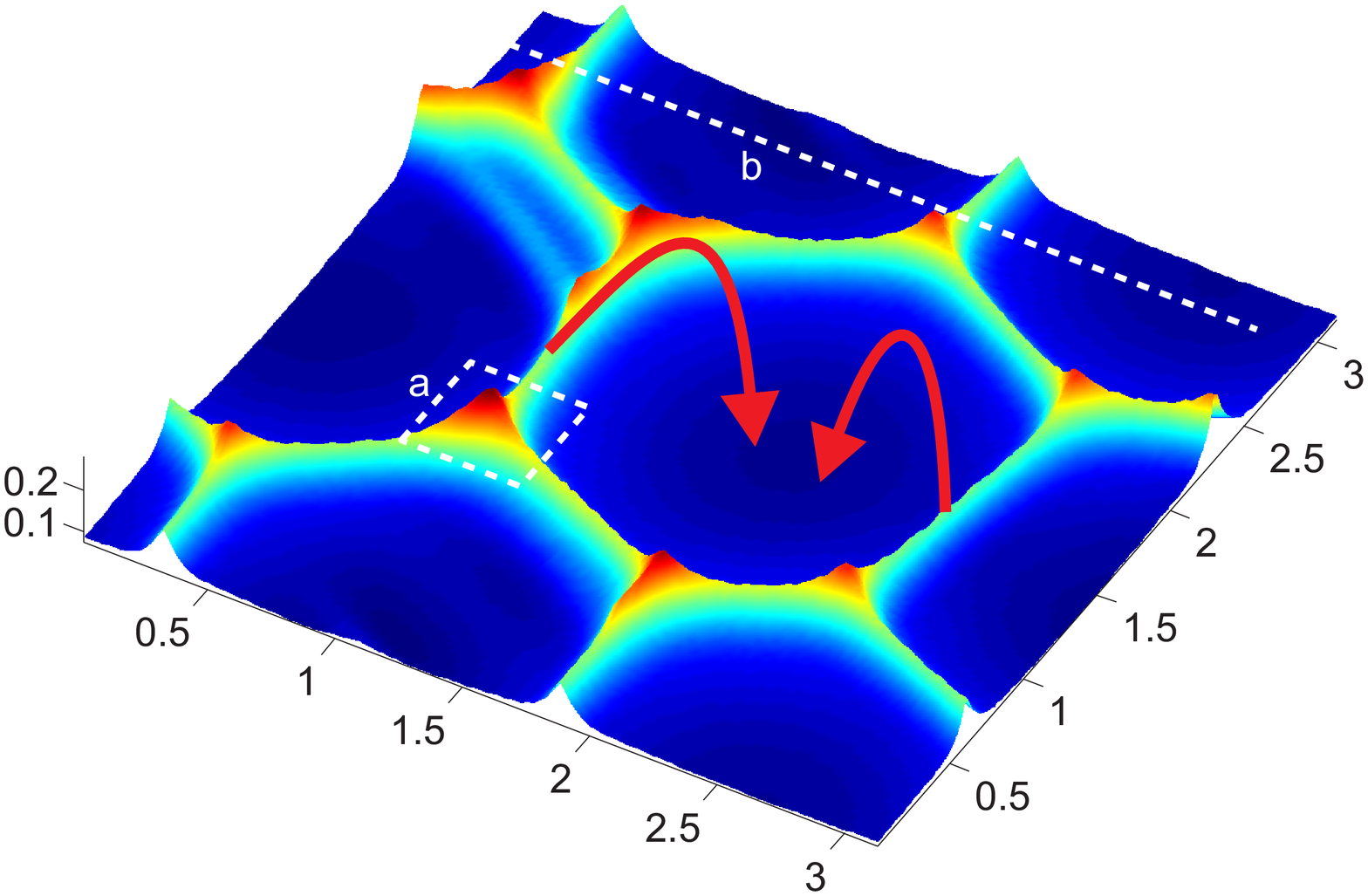}
\includegraphics[width=.39\textwidth]{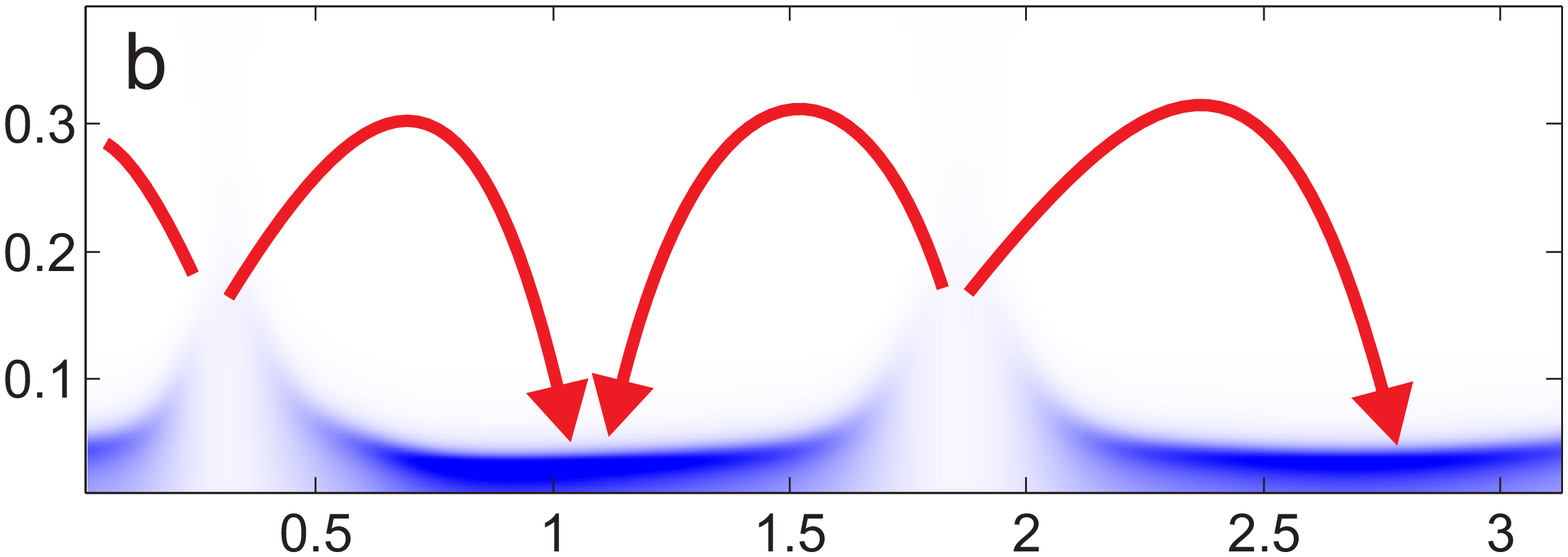}
\begin{picture}(0,0)
\put(-190,240){$j(x,z)$}
\put(-190,20){$y_m(x,z)$}
\put(0,20){$\rho(x,y,z_*)$}
\put(90,120){\footnotesize a}
\put(100,90){\includegraphics[width=.1\textwidth]{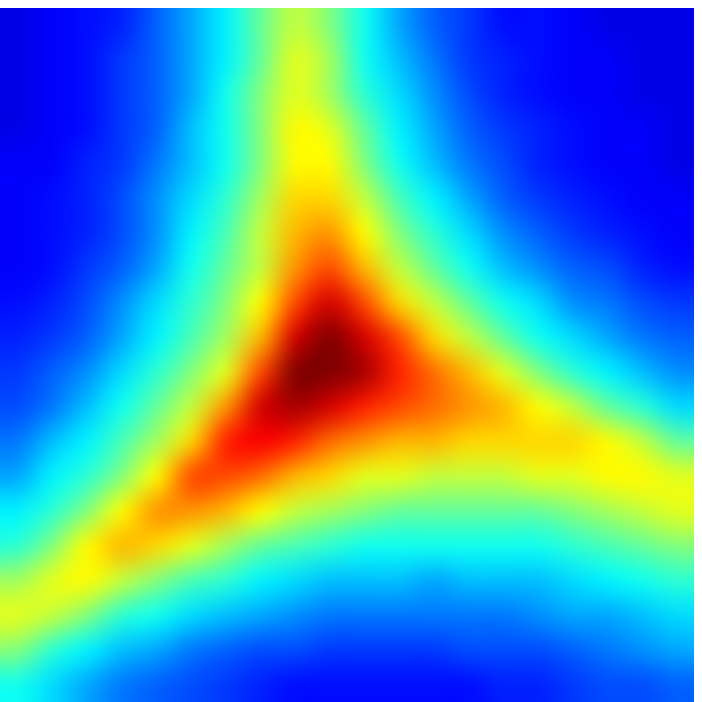}}
\end{picture}
\caption{(Color online) Hexagonal coherent structures, $\Delta V=50$,
$\varkappa=0.1$, and $\nu=10^{-3}$. Profile of ESC region, $y_m(x,z)$;
(a)~vicinity of the pyramid top; (b)~cross-section $z=z_*$ of the charge
density~$\rho$. Distribution of the electric current $j(x,z)$ on the membrane
surface. The arrows schematically show the direction of the liquid flow.}
\label{fig:hexagon}
\end{figure}

\begin{figure}[hbtp]
\raggedright
\includegraphics[origin=rb,angle=270,width=.6\textwidth]{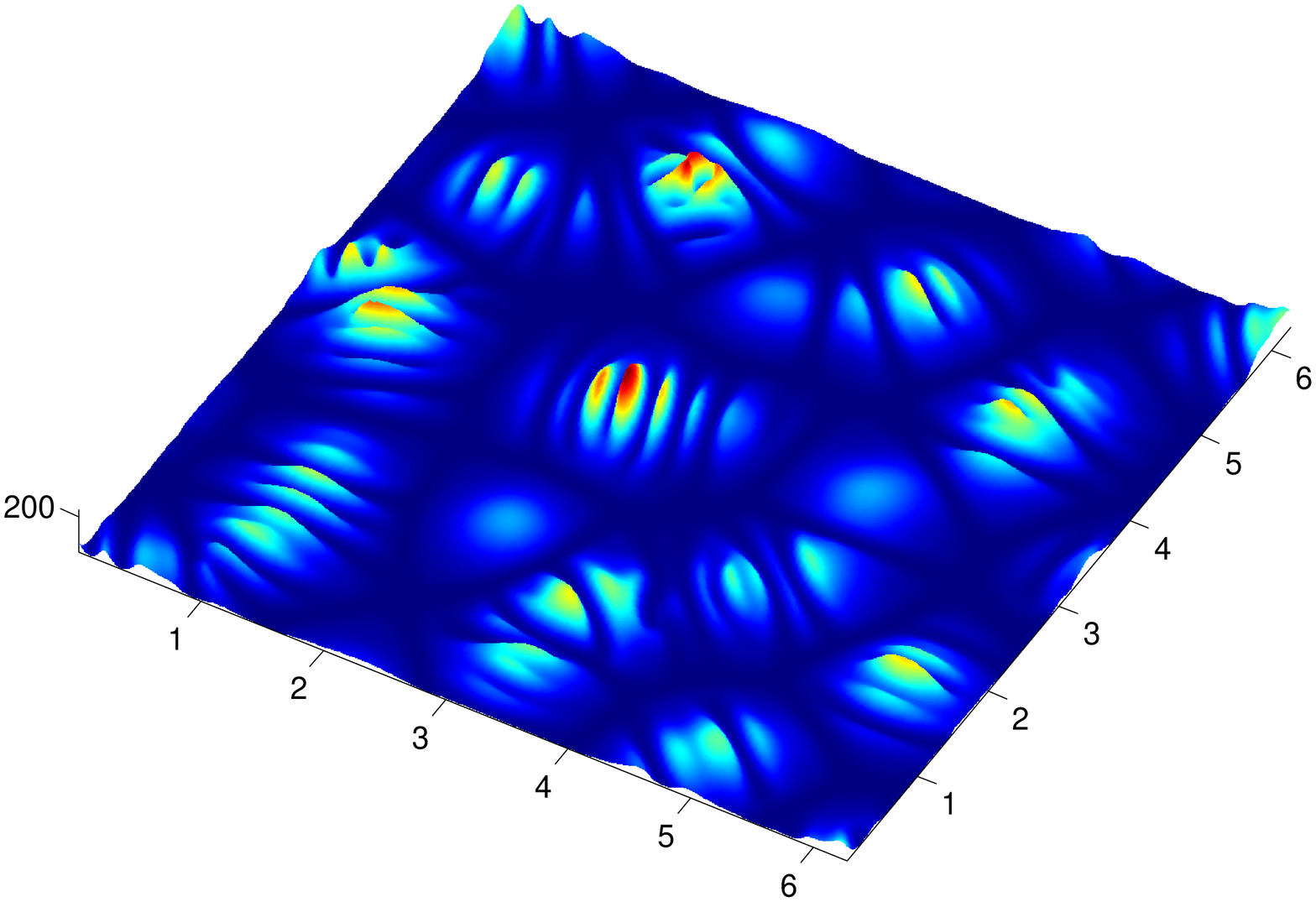}

\centering
\includegraphics[origin=rb,angle=270,width=.6\textwidth]{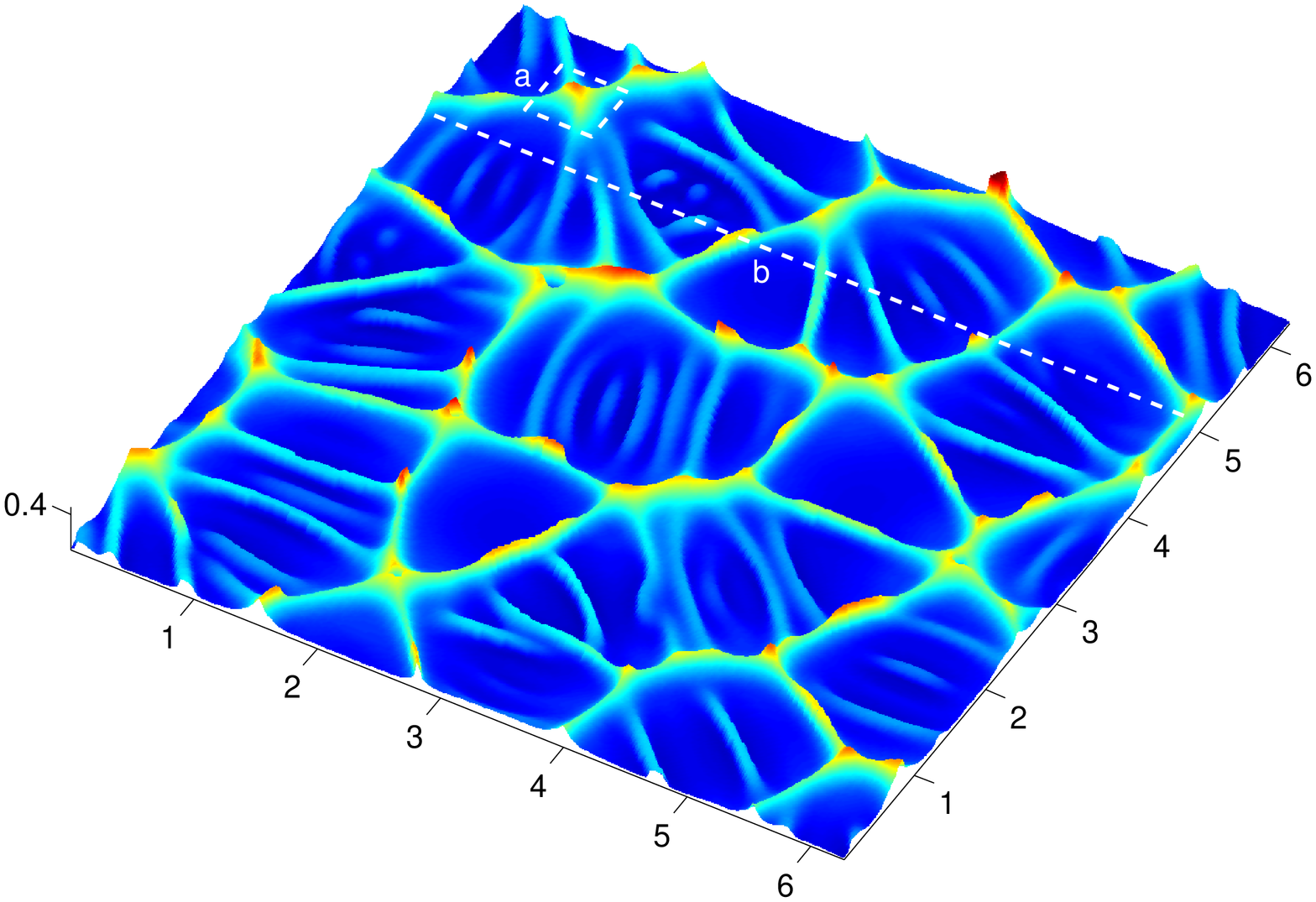}
\includegraphics[width=.39\textwidth]{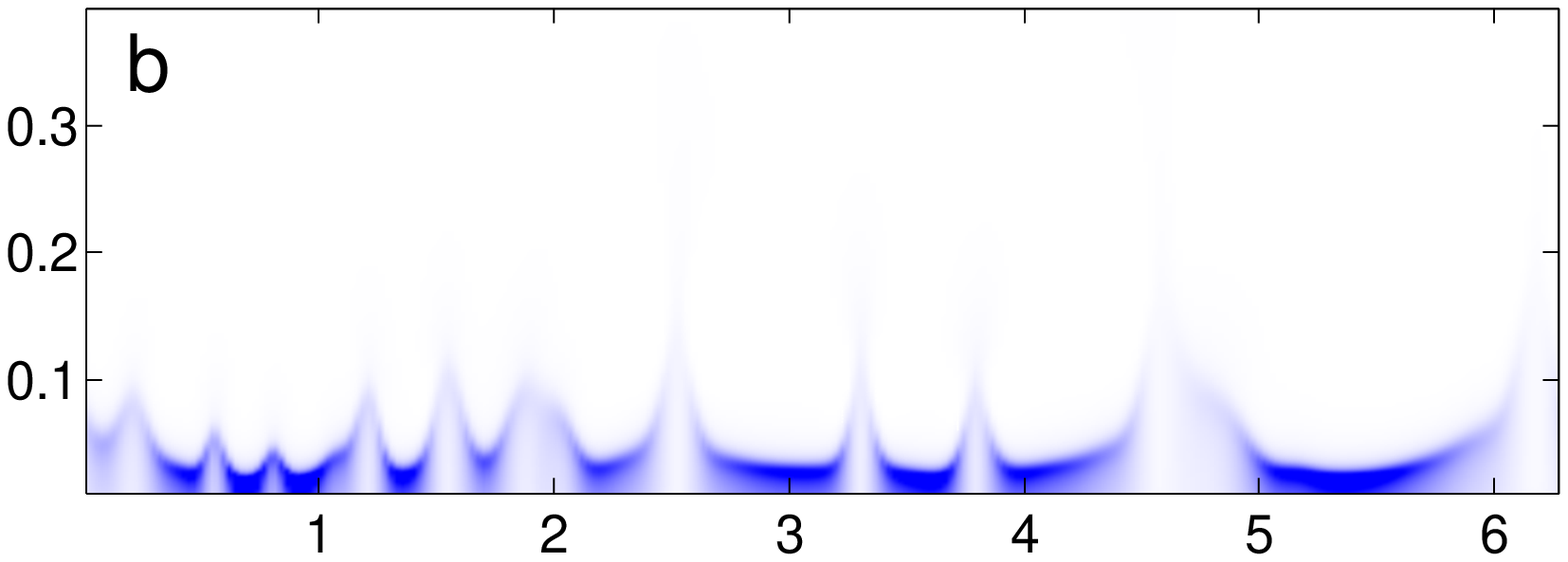}
\begin{picture}(0,0)
\put(-190,250){$j(x,z)$}
\put(-190,30){$y_m(x,z)$}
\put(0,30){$\rho(x,y,z_*)$}
\put(-25,190){2D solitary pulses}\thicklines
\put(-25,185){\vector(-4,-1){37}}
\put(-25,185){\vector(-1,-1){56}}
\put(-25,185){\vector(-1,-3){33}}
\put(90,120){\footnotesize a}
\put(100,90){\includegraphics[width=.1\textwidth]{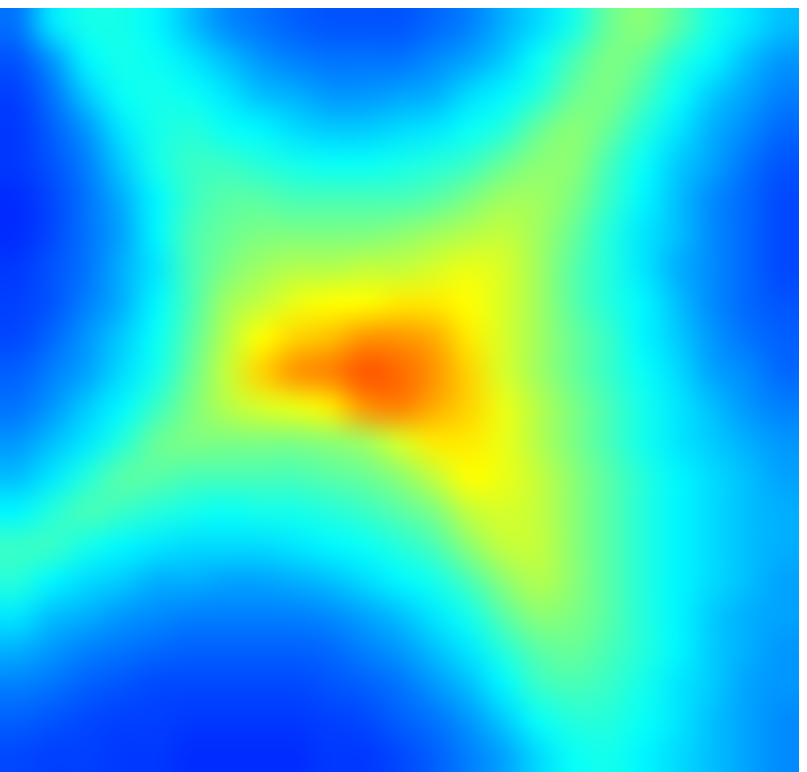}}
\end{picture}
\caption{(Color online) Snapshot of $y_m$ for the spatiotemporal chaos;
(a)~vicinity of the pyramid top; (b)~cross-section $z=z_*$ of the charge
density~$\rho$. Distribution of the electric current $j(x,z)$ on the membrane
surface. $\Delta V=60$, $\varkappa=0.1$, and $\nu=10^{-3}$.}
\label{fig:chaos}
\end{figure}

Let us consider some important details of these characteristic patterns: the
electroconvective rolls, the hexagons, and the spatiotemporal chaos. In order to
present a full picture of the behavior, it is instructive to analyze together
the distribution of $y_m(x,z)$ along the membrane surface, the charge density
$\rho$ inside the ESC region, and the electric current $j(x,z)$ determined by
Eq. \eqref{eq777}. Their typical snapshots are depicted in \figref{fig:2d},
\figref{fig:hexagon}, and \figref{fig:chaos}.

For rolls, the profile $y_m(x,z)$ is shown in \figref{fig:2d}; it has long flat
and short wedge-like regions with a cusp at the top. The wedge angle or the
angle between the wedge faces is rather conservative, it practically does not
depend on the parameters and is about $105^\circ \div 118^\circ$.

After loss of stability and after the corresponding bifurcation, the rolls turn
into steady three-dimensional structures, see \figref{fig:hexagon}; these
structures are conventionally called ``hexagons.'' The $y_m(x,z)$ profile
consists of six wedge-like lateral faces and six pyramids are located at their
intersection. In the remaining area, $y_m(x,z)$ is flat and situated in the
lowlands. It is interesting that the angle of wedge-like faces is close to that
for the 2D rolls, and the dependence of this angle on the parameters of the
problem is also weak. A rough evaluation of the pyramid angle gives its value as
about $85^\circ \div 90^\circ$.

Snapshots of the spatiotemporal chaos are shown in \figref{fig:chaos}. Now, the
$y_m(x,z)$ distribution consists of a combination of triangles, quadrangles, and
hexagons whose location and form change chaotically. The sides of these
geometrical figures are wedges with an angle averaged over time of about
$110^\circ$. The pyramids formed at the intersection of the sides have a
time-average angle at the top of about $90^\circ$.

The numerical resolution of the charge density $\rho$ in the thin ESC layer is
shown for our three basic patterns in the left part of \figref{fig:2d},
\figref{fig:hexagon}, and \figref{fig:chaos}. The darker regions correspond to
large charge densities $\rho$ with a rather sharp boundary between the ESC
region, $0<y<y_m$, and the diffusion region, $y>y_m$. The portions with a small
charge in the spikes are joined by the flat regions of large charge. For all
three regimes, the minimum of $y_m$ corresponds to the maximum of the charge
density. At the top of the pyramids, where $y_m(x,y)$ reaches its maximum
maximorum, the $\rho$ distribution always has its minimum minimorum.
 
The electric current at the membrane surface, $j(x,z)$, is another important
characteristic value whose description complements our understanding of the
system behavior, see the top of \figref{fig:2d}, \figref{fig:hexagon}, and
\figref{fig:chaos}. For all three basic coherent structures, $j(x,z)$
qualitatively replicates the $y_m(x,z)$ profile and $\rho$ distribution in the
ESC layer, but smooths their sharp details: for 2D rolls, the localized
wedge-like profile of $y_m(x,z)$ turns into the nearly sinusoidal profile of
$j(x,z)$; for the 3D regular patterns, the hexagon turns into a circle; the
triangles, quadrangles, and hexagons of the spatiotemporal chaos transform into
a system of circles and ellipses. Moreover, the electric current $j(x,z)$ has
minimal values in the vicinity of the cusps and is maximal in the flat regions
of the $y_m$ and $\rho$ distributions. We attribute this behavior to the fact
that the electrical conductivity is smaller in the cusp regions and larger in
the lowlands. 

Our simulations show that liquid always flows upwards from the cusp points of
the $y_m$ and $\rho$ distributions and returns to the membrane surface moving
towards the flat regions of the $\rho$ distribution. An array of vortex pairs is
formed; it is schematically shown in the figures by the arrows. The
characteristic size of the electroconvective rolls varies within the range
$1.3 \div 2.0$, of the regular hexagonal structures the size is about $1.5$, and
for the spatiotemporal chaos, about $1.2$.

An interesting phenomenon found is the generation of two-dimensional running
solitary waves (pulses). Such waves form spontaneously either inside the
hexagonal structure or at one of its lateral sides with subsequent propagation
towards the opposite side of the hexagon, see \figref{fig:chaos}. For relatively
small drives $\Delta V-\Delta V_*$, the pulse generation is a rare event,
moreover the pulse decays during its propagation and eventually completely
disappears. As $\Delta V-\Delta V_*$ increases, this generation occurs more
frequently, the pulse amplitude increases, and the pulse can propagate without
decaying and reach the opposite side of the hexagon. If at the neighboring side
or somewhere else another solitary wave forms and then departs, a complex
pulse--pulse interaction occurs; depending on the spatial location of the waves,
it can be a head-on or an oblique interaction. For large $\Delta V-\Delta V_*$,
the pulse--pulse interaction becomes strong enough to destroy the hexagonal
structure and a transition to spatiotemporal chaos results from the interaction.
Note that similar phenomena of pulse generation and pulse--pulse interactions
have been observed for other kinds of instability, Marangoni--B\'enard
convection \cite{Vel1,Vel2} and falling liquid films \cite{ChD}.

\section*{Conclusions}

A~direct numerical simulation of the elektrokinetic instability in its
three-dimensional formulation was carried out. A~special numerical algorithm was
developed. The calculations employed parallel computing. Three characteristic
patterns, which correspond to the overlimiting currents, were observed:
two-dimensional electroconvective rolls, three-dimensional regular hexagonal
structures, and three-dimensional structures of spatiotemporal chaos, which are
combinations of unsteady hexagons, quadrangles and triangles. The distinguishing
features of the regular and chaotic three-dimensional regimes were found. The
transition from the steady regular three-dimensional patterns to the
spatiotemporal chaos was found to be accompanied by the generation of
interacting two-dimensional solitary pulses.

\section*{Acknowledgements}

This work was supported, in part, by the Russian Foundation for Basic Research
(Projects No. 12-08-00924-a, 13-08-96536-r\_yug\_a).

\end{document}